\documentclass{desyproc}
\usepackage{sidecap}
\begin{document}
%------------------------------------
\title{Quarkonium dissociation in a thermal medium}

%for single authors the superscripts are optional
\author{{\slshape  Jakub Jankowski and David Blaschke}\\[1ex]
Institute for Theoretical Physics, University of Wroclaw, 
pl Maxa Borna 9, 50-204 Wroclaw, Poland  }

% if the proceedings are available online (e.g. at Indico)
% please enter the contribution ID or file_name below for the DOI
%\contribID{32}
\contribID{jankowski\_jakub}

% TO THE CONFERENCE EDITORS: 
% please update the following information      
% before sending the template to the authors
% \confID{800}  % if the conference is on Indico uncomment this line
\desyproc{DESY-PROC-2008-xx}
\acronym{HQP08}%if you want the Acronym in the page footer uncomment this line
\doi  % if there is an online version we will register DOIs

\maketitle

\begin{abstract}
We investigate the Mott effect for heavy quarkonia due to Debye screening 
of the heavy quark potential in a plasma of massless quarks and antiquarks. 
The influence of residual color correlation is investigated by coupling the 
light quark sector to a temporal gauge field driven by the Polyakov loop 
potential. This leads to an increase of the Mott dissociation temperatures 
for quarkonia states which stabilizes in particular the excited states, but
has marginal effect on the ground states.
\end{abstract}

%%%%%%%%%%%%%
%%%%%%%%%%%%%

\section{Introduction}

Since the suggestion of $J/\psi$ suppression as a signal of quark-gluon plasma
(QGP) formation by Matsui and Satz \cite{Matsui:1986dk} in 1986 the problem of
quarkonium dissociation in hot and dense strongly interacting matter has played
a key role for QGP diagnostics in relativistic heavy-ion collision experiments.
The original idea was that in a QGP the string tension of the confining 
potential vanishes and the residual one-gluon exchange interaction undergoes 
a Debye screening by the color charges of the plasma. 
When the temperature dependent Debye radius $r_D(T)$ (the inverse of the Debye 
mass $m_D(T)$) becomes shorter than the Bohr radius of the charmonium ground 
state ($J/\psi$) then the Mott effect \cite{Mott:1968zz} (bound state 
dissociation) occurs and the corresponding temperature is 
$T_{\rm Mott}^{J/\psi}$. 
This simple idea grew up to a multifacetted research direction when not only in
the first light ion - nucleus collisions at the CERN NA38 experiment, but also
in proton - nucleus collisions at Fermilab $J/\psi$ suppression has been found
so that there is not only a QGP but also a cold nuclear matter effect on 
charmonium production, see \cite{Rapp:2008tf} for a recent review.
%In the present contribution, however, we want to focus on simple analytical
%estimates of the Mott dissociation temperatures for quarkonia states in a 
%QGP, by restricting ourselves to the evaluation of the Debye mass in a quark
%plasma in the chiral limit, and treating gluons as a homogeneous background
%field.

If one wants to explore the question of screening in a plasma more in detail 
then a variety of appraoches is available in the literature, from the original 
Debye-H\"{u}ckel approach \cite{Dixit:1989vq} where one can study any vacuum 
potential (for example the Cornell potential) and see its medium modification,
over the thermodynamic Green functions approach to the ab-initio studies of
heavy-quark potentials in lattice QCD. 
With the obtained medium-dependent potentials one can then study the bound 
state problem by solving the nonrelativistic Schr\"{o}dinger equation or, more
systematically, the thermodynamic $T$ - matrix for quarkonia 
\cite{Cabrera:2006wh}. 

On the other hand one may calculate proper correlators directly from lattice 
QCD and extract from them spectral functions \cite{Asakawa:2000tr}. 
There is an intriguing disagreement between the Mott temperatures deduced from
these spectral functions and those of the potential models: 
the latter are much smaller than the former! 
From the lattice data for quarkonium correlators one has extracted 
$ T^{\rm Mott}_{J/\psi} \approx 1.9 T_c $  while in 
potential model calculations $ T^{\rm Mott}_{J/\psi} \approx 1.2 T_c$. 
This problem has lead to the discussion of the proper thermodynamical function 
to be used as a potential in the Schr\"odinger equation. 
Should it be the free energy or the internal energy? 
We will not follow this question in the present work, but refer to 
\cite{Rapp:2008tf,Satz:2005hx} and references therein. 

Here we examine a simple model of screening as derived from one-loop 
calculations in thermal quantum field theory, and make a small extension of 
this result by putting the internal fermion lines in a constant temporal 
gauge field which mimics confining gluon dynamics (Polyakov-loop potential).
In our approach the medium is made of plasma of massless quarks, described by
the chirally symmetric phase of the Nambu--Jona-Lasinio (NJL) model of QCD  
\cite{Klevansky:1992qe}. 
Confinement is implemented in the most simple way by coupling the system 
to the Poyakov loop variable - resulting in the so called Polakov loop 
NJL-model (or PNJL model). Recently it has been successfully used to
reproduce lattice data \cite{Ratti:2005jh} or to describe light meson physics 
at finite temperatures and densities \cite{Hansen:2006ee}.

%%%%%%%%%%%%%%%%%
%%%%%%%%%%%%%%%%%

\section{Debye-screening in a PNJL quark plasma}

Given the self energy (polarization function) of a boson that 
mediates the interaction, the screened potential is given by a resummation of 
one-particle irreducible diagrams ("bubble" resummation = RPA) \cite{LeBellac}
\begin{equation}
V_{\rm sc}(q) = {V(q)}/[{1 + F(0;{\bf q})/q^{2}}]~, 
\label{Vsc}
\end{equation}
where we take  
$ V(q) = -\frac{4}{3}{g^{2}}/{q^2}$, $ q^{2} = |\bf{q}|^{2} $
as the unscreened vacuum potential. 
The longitudinal gluon polarization function $ F(0; {\bf q}) = - \Pi_{00}(0; {\bf q}) $ in the finite $ T $ case can be calculated within 
perturbative thermal field theory where it takes the form
\begin{equation}
\Pi_{00}(i\omega_{l};{\bf q} ) 
= Tg^{2}\sum_{n=-\infty}^{\infty} \int\frac{d^{3}p}{(2\pi)^{3}} 
{\textrm{ Tr}} [\gamma^{0}S_{\Phi}(i\omega_{n}; {\bf p})
\gamma^{0}S_{\Phi}(i\omega_{n}-i\omega_{l}; {\bf p} - {\bf q})]~,
\end{equation}
where $\omega_{l}=2\pi lT$ are the bosonic and $\omega_{n}=(2n+1)\pi T$
are the fermionic Matsubara frequencies of the imaginary-time formalism.
The symbol ${\textrm{ Tr}}$ stands for traces in color, flavor and Dirac 
spaces.
$S_{\Phi}$ is the propagator of a massless fermion coupled to the homogeneous 
static gluon background field $\varphi_3$. Its inverse is given by 
\cite{Ratti:2005jh,Hansen:2006ee}
\begin{equation}
S^{-1}_{\Phi}( {\bf p}; \omega_{n} ) = 
{\bf \gamma\cdot p} + \gamma_{0}i\omega_{n} -\lambda_{3}\varphi_3~,
\label{coupling}
\end{equation}
where $\varphi_3$ is related to the Polyakov loop variable defined by
\cite{Ratti:2005jh}
$$ \Phi(T) = \frac{1}{3}\rm Tr_c (e^{i\beta\lambda_{3}\varphi_{3}}) 
= \frac{1}{3}(1 + 2\cos(\beta\varphi_3) )~. $$ 
The physics of $\Phi(T)$ is governed by the temperature-dependent Polyakov 
loop potential ${\cal{U}}(\Phi)$, which is fitted to describe the lattice data 
for the pressure of the pure glue system  \cite{Ratti:2005jh}. 
After performing the color-, flavor- and Dirac traces and making the fermionic 
Matsubara summation, we obtain in the static, long wavelength limit 
\begin{equation}
\Pi_{00}(0; {\bf q} )  
= \frac{2N_{\rm dof}g^{2}}{\pi^2} \int_{0}^{\infty}dp\, 
p^{2}\frac{\partial f_\Phi}{\partial p} 
= - \frac{ 4N_{\rm dof} g^{2}}{\pi^2} \int_{0}^{\infty}dp\,p f_{\Phi}(p) 
= -\frac{N_{\rm dof} g^{2}T^{2}}{3}I(\Phi) = - m_{D}^2(T) ~.
\label{debyemass}
\end{equation}
where $ m_D(T)$ is the Debye mass, the number of degrees of freedom 
is $N_{\rm dof}=N_c\ N_f=6$ and $f_\Phi(p)$ is quark distribution 
function \cite{Hansen:2006ee}. The screened potential is thus
\begin{equation}
\label{Vs}
V_{\rm sc}(q) = -4\pi \alpha/[q^2+m_D^2(T)]~.
\end{equation}
In comparison to the free fermion case \cite{LeBellac,Beraudo:2007ky} the 
coupling to the Polyakov loop variable $\Phi(T)$ gives rise to a modification 
of the Debye mass, given by the integral
\begin{equation}
I(\Phi) = \frac{12}{\pi^2}\int_{0}^{\infty}\,
dx\,x\frac{\Phi(1+2e^{- x})e^{- x}+e^{-3 x}}
{1 + 3\Phi(1 + e^{- x})e^{- x}+e^{-3 x}}.
\end{equation}
In the limit of deconfinement ($\Phi = 1$), the case of a massless
quark gas is obtained ($I(1)=1$) while for confinement ($\Phi = 0$) one finds
that $I(0)=1/9$. 
For the temperature dependence of $\Phi(T)$ we employ in the following 
chapter the results of a nonlocal PNJL model \cite{Blaschke:2007np}.

%%%%%%%%%%%%
%%%%%%%%%%%%

\section{Variational ansatz and estimation of Mott temperatures} 

Here we will use the derived potential in the quantum mechanical way to 
estimate the dissociation temperature. With the trial wave function for the 
1S state
\begin{equation}
 \psi_{\rm 1S}(r;\gamma) = \sqrt{\frac{\gamma^{3}}{\pi}}\exp(-\gamma r) 
\end{equation}
and the non-relativistic two-body Hamiltonian
\begin{equation}
H= -\frac{\nabla^{2}}{m_{Q}}-\frac{\alpha}{r}e^{-m_{D}(T)r}  ~, 
%\nonumber
\end{equation}
where the potential term is the Fourier transform of screened Coulomb 
potential (\ref{Vs}),
we obtain the energy functional for the Ritz variational principle 
\begin{equation}
E_{\rm 1S}(\gamma,T)= 
\langle\psi_{\rm 1S}(\gamma)\mid H\mid \psi_{\rm 1S}(\gamma)\rangle 
=\frac{\gamma^2}{m_Q}-\frac{4\alpha\gamma^3}{(m_D(T) + 2\gamma)^2}~.
%\nonumber
\end{equation}
Simultaneously satisfying the conditions for the ground state energy 
${dE_{\rm 1S}(\gamma,T)}/{d\gamma}=0$
and for a vanishing binding energy (Mott effect),
$E_{\rm 1S}(\gamma,T^{\rm Mott})=0$,
%
%\begin{equation}
%\frac{dE_{\rm 1S}(\gamma)}{d\gamma}\bigg|_{T^{\rm Mott}} = 0 ~~
%\textrm{and} ~~
%E(\gamma)\big|_{T^{\rm Mott}}= 0~,
%\label{condition}
%\end{equation}
provides us with an analytic expression for the critical Debye mass 
%
%\begin{equation}
$m_D(T^{\rm Mott}_{\rm 1S}) = 2\gamma~.$
%\label{MottCondition}
%\end{equation}
%
Using once again the condition $E_{\rm 1S}(\gamma,T^{\rm Mott}) = 0 $,
results in the Mott condition for the Debye potential \cite{Mott:1968zz}
\begin{equation}
r_{D}( T^{\rm Mott}_{\rm 1S})= a_{0}~,
\label{MottCondition}
\end{equation}
where $ a_{0} = 2/ (\alpha m_{Q}) = 1/\sqrt{\varepsilon_0 m_Q} $ is the Bohr 
radius and $\varepsilon_0=\alpha^2 m_Q/4$ the binding energy of ground state 
in the vacuum ($m_D=0$).
%We have used the relation $r_D=1/m_D$ between Debye radius and Debye mass.
%
Due to the temperature dependence of the Debye mass, we 
obtain the Mott dissociation temperature in the massless quark gas 
(for $ \Phi = 1 $)
\begin{equation}
T^{\rm Mott} = \sqrt{{3 \varepsilon_0 m_Q}/{N_{\rm dof}}}/{g}
= \sqrt{{\sqrt{\varepsilon_0 m_Q^3}}/{(2\pi N_{\rm dof})}}~.
\label{dissT}
\end{equation}
%%%%%%%%%%%%%%%%%
\begin{SCtable}[3.0]
\caption{Mott temperatures $ T^{\rm Mott}$ ($ T^{{\rm Mott},\Phi}$) 
%for dissociation of heavy quarkonia ground states 
according to Eq. (\ref{dissT}) (Eq. (\ref{dissTPhi})) for a massless ideal 
quark gas (PNJL model). 
The critical temperature %for the phase transition 
is $T_c=202$ MeV \cite{Blaschke:2007np}.
The parameters are fixed to reproduce quarkonium states in vacuum as 
Coulombic bound states \cite{Arleo:2001zj}. 
In the charmonium (bottomonium) system the heavy quark mass $m_Q$ is 
$ m_{c}=1.94$ GeV ($m_b=5.1$ GeV) and the ground state binding energy 
$\varepsilon_0$ is $0.78$ GeV ($0.75$ GeV).
}
\begin{tabular}{|c||c|c|}
\hline
& $ T^{\rm Mott}/T_c$ &$T^{\rm Mott,\Phi}/T_c$\\
\hline
\hline
$J/\psi$	&$  1.25$ & $1.37$  \\
$\chi_c$	&$  0.83$ & $1.11$  \\
$\psi'$	&$  0.66$ & $0.99$  \\
\hline
$\Upsilon$ 	   &$  2.50$ & $2.50$  \\
$\chi_b$           &$  1.72$ & $1.73$  \\
$\Upsilon'$   &$  1.28$ & $1.39$  \\
\hline
\end{tabular}

\label{tab:limits}
\end{SCtable}
In Tab.~\ref{tab:limits} we give the parameter values according to set (i)
of Ref. \cite{Arleo:2001zj}.
The Poyakov loop variable contribution considered in previous section affects the Mott temperature in Eq. (\ref{MottCondition}) 
and gives the following formula
\begin{equation}
\label{dissTPhi} 
T^{{\rm Mott},\Phi} = T^{\rm Mott}/\sqrt{I(\Phi)}~.
\end{equation}
This means that in the case of confining color correlations ($0\le\Phi<1$) 
the Debye screening radius is larger than in a free quark gas at the same 
temperature, so that bound states get stabilized against thermal dissociation
by color screening. 
\begin{figure}[!ht]
\centerline{
\includegraphics[width=0.65\textwidth,height=0.6\textwidth]{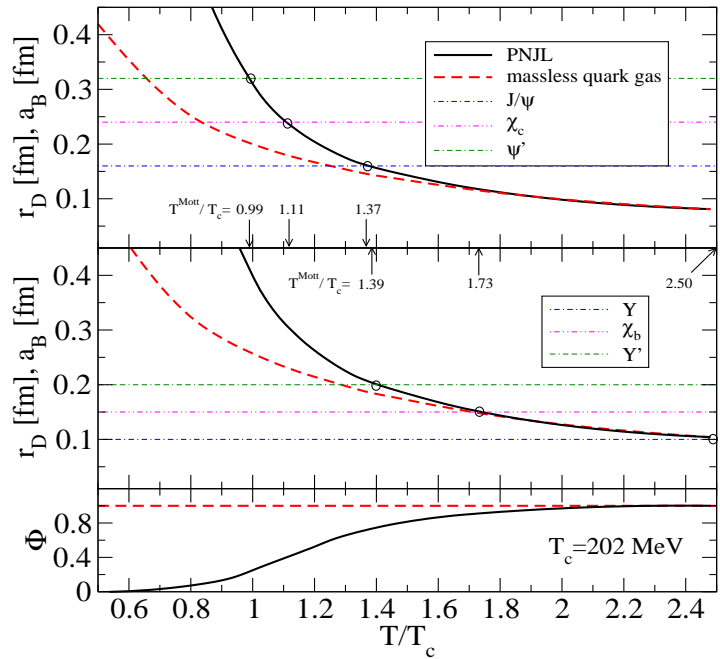}}
\caption{Temperature dependence of the Debye screening radius $r_D(T)$ for 
charmonium (upper panel) and bottomonium (middle panel) with (solid lines) and 
without (dashed lines) coupling to the Polyakov loop $\Phi(T)$ shown in the 
lower panel (from \cite{Blaschke:2007np}). 
Sequential dissociation of quarkonia states occurs at the Mott
temperatures $T^{\rm Mott}$ for which their Bohr radius equals $r_D(T)$.}
\label{Fig:Debye}
\end{figure}

%%%%%%%%%%%%%%%%%%%%%%%%%%%%%%
The influence of the Poyakov loop variable on the dissociation temperature is 
summarized in 
Figure \ref{Fig:Debye} which shows the temperature dependence of the Debye 
radius $ r_D(T) = {1}/{m_D(T)}$ compared to the Bohr radii of the 
low-lying states of the charmonium and bottomonium family, respectively.
Due to its larger mass and smaller Bohr radius, the $\Upsilon$ dissociates at 
higher temperatures than the $J/\psi$, where the free quark gas case is 
almost reached and the Mott temperatures which almost coincide for the cases 
with and without coupling to the Polyakov loop field. 
For the lighter $J/\psi$ there is a noticeable stabilization due to the 
coupling to the Polyakov loop potential which results in an increase of the 
Mott dissociation temperature from $1.25~T_c$ to $1.37~T_c$, still more similar
to results of nonrelativistic potential models rather than the still higher 
dissociation temperatures conjectured from the spectral functions deduced from
lattice data for heavy quarkonium correlators by the maximum entropy method. 
For estimating the Mott temperatures of the excited quarkonia states we have 
employed the scaling of bound state radii: $r_{\chi_c}=1.5~r_{J/\psi}$,  
$r_{\psi'}=2.0~r_{J/\psi}$ as obtained in the Cornell-type potential model
\cite{Satz:2005hx}.

%%%%%%%%%%%%%%%%%%%%%%%%%%%%%%

\section{Conclusions}

We have applied the methods of thermal field theory to estimate the effects
of Debye screening on heavy quarkonia bound state formation. In order to
account for residual effects of confining color correlations in the deconfined
phase, we have used the PNJL model in the evaluation of the 
one-loop polarization function. As expected, a stabilization of bound states
in the vicinity of the critical temperature for $T>T_c$ is obtained.
We applied Ritz' variational principle to derive the Mott criterion for bound 
states of the statically screened Debye potential and obtained Mott 
temperatures in good agreement 
with previous results from nonrelativistic potential models exploiting 
lattice QCD singlet free energies as potentials in the Schr\"odinger equation 
for heavy quarkonia. 
This agreement with previous results (see, e.g., Ref. \cite{Satz:2005hx}) 
includes also higher quarkonia resonances for which the 
stabilization effects is more pronounced.

%%%%%%%%%%%%%%%%%
%%%%%%%%%%%%%%%%%

\section{Acknowledgments}

JJ received financial support from the Bogoliubov-Infeld program for his 
participation in the Helmholtz International Summer Schools in Dubna. 
The work of DB has been supported in part by the Polish Ministry for Science
and Higher Education under grant No. N N 202 0953 33 and by RFBR grant No.
08-02-01003-a. 
%\section{Bibliography}

% ****************************************************************************
% BIBLIOGRAPHY AREA
% ****************************************************************************

\begin{footnotesize}
% IF YOU DO NOT USE BIBTEX, USE THE FOLLOWING SAMPLE SCHEME FOR THE REFERENCES
% ----------------------------------------------------------------------------

% ----------------------------------------------------------------------------

% IF YOU USE BIBTEX,
% - DELETE THE TEXT BETWEEN THE TWO ABOVE DASHED LINES
% - UNCOMMENT THE NEXT TWO LINES AND REPLACE 'Name_Of_Your_BibFile'

%\bibliographystyle{unsrt}
%\bibliography{Name_Of_Your_BibFile}
% example of Name_Of_Your_BibFile.bib
% @Article{Turcato:2006ch,
%      author    = "Turcato, M.",
%  collaboration = "ZEUS and H1",
%      title     = "Lepton flavour violation and charmonium physics at HERA",
%      journal   = "Nucl. Phys. Proc. Suppl.",
%      volume    = "162",
%      year      = "2006", 
%      pages     = "283-287",
%      SLACcitation  = "%%CITATION = NUPHZ,162,283;%%"
% }
% 
% @Unpublished{Gogitidze:2007du,
%      author    = "Gogitidze, N.",
%  collaboration = "H1", 
%      title     = "Prompt photons and particle momentum distributions at
%                   HERA", 
%      year      = "2007",
%      note    = "hep-ex/0701033",
%      SLACcitation  = "%%CITATION = HEP-EX 0701033;%%"
% }

\end{footnotesize}

% ****************************************************************************
% END OF BIBLIOGRAPHY AREA
% ****************************************************************************

\end{document}